\documentstyle{aipproc}
\newcommand{\LI}{\hbox to\hsize}

\newcommand{\BEQ}{\begin{equation}}
\newcommand{\EEQ}{\end{equation}}

\newcommand{\DEG}[1]{\mbox{$ #1^{\rm o}$}}

\newcommand{\incircle}[1]{\mbox{{\hbox{$\bigcirc$}\kern-0.7em
\lower0.05ex\hbox{\mbox{{\scriptsize\rm #1}}}}}}
\newcommand{\CF}{\mbox{\em cf.\/ }}

\newcommand{\ETAL}{\mbox{\em et. al.\/ }}

\begin{document}
\input epsf
\title{Observation of UHE Neutrino Interactions from Outer Space}
\author{G. Domokos and S. Kovesi-Domokos}
\address{Department of Physics and Astronomy\\
The Johns Hopkins University\\
Baltimore, MD~21218\thanks{E-mail: skd@haar.pha.jhu.edu}}
\maketitle
\begin{abstract}
The interaction of UHE neutrinos can be observed from outer space. 
The advantage of the proposed method is that the Earth can be used as an 
energy filter as well as a target.
We sketch the potentials of these observations in searching for particle 
physics beyond the Standard Model of elementary particle interactions 
as well as observations of astronomical objects such as active galactic nuclei.
\end{abstract}
\section{Introduction}
It has been recognized that the observation of UHE neutrino interactions 
can serve a dual purpose:
\begin{itemize}
\item The observation of objects of high activity and high column density
in the sky,
(a typical example is an AGN) which otherwise are inaccessible to observation
using electromagnetic waves at almost any wavelength;
\item a search for particle physics not described by the current ``Standard
Model'' of elementary particle interactions.
\end{itemize}

The first entry on our list has been widely explored in a number of
articles and it has been discussed at a variety of conferences; the
first meeting with the purpose of exploring all the astrophysical uses
of neutrino telescopy has been the Hawaii conference\cite{hawaii},
followed by a series of meetings on neutrino astronomy in 
Venice\cite{venice} and in Pylos\cite{nestor}.

The particle physics aspect of neutrino telescopy is somewhat less known; 
however, in our opinion, it is equally important. A reasonably comprehensive
discussion of the particle physics uses of neutrino telescopy is 
contained in an article published in one of the Venice
proceedings\cite{nestor}.

In this paper we report the results of a first exploration of the particle
physics uses of a novel type of neutrino detectors:  the use of 
orbiting detectors, such as the {\em O}rbiting {\em W}ide angle {\em L}ight
collector (OWL) as well as a detector flown on a space station.

This paper is organized as follows. In the next section we outline the
potential advantages of an orbiting detector used as a neutrino telescope.
Section~3 contains a brief description of our first results. The final section
contains a discussion and outlook.

Most of the results reported here are preliminary in nature: they will be 
followed up by more detailed investigations. Nevertheless, the ones reported
here look exciting and, for that reason, worth discussing.
\section{``Eye in the Sky''}
Conventional (?) neutrino telescopes, such as AMANDA, BAIKAL, 
SUPER-KAMIOKANDE and the
future NESTOR have the advantage that they are based on technologies
which, by now, are reasonably well explored. Their basic feature is
that they are capable of a good directional resolution (currently
of the order of \DEG{1} or so. (This is about the resolution of 
Galileo's telescopes: neutrino telescopy is at the age optical
telescopy was in Galileo's time\ldots) These telescopes therefore 
provide a valuable tool for exploring the neutrino sky for both
point like (AGN?) and diffuse sources, such as atmospheric neutrinos and
neutrinos emitted by baryonic dark matter interacting with the
primary cosmic radiation\cite{bruceelliott}. Likewise, on the
particle physics side, such detectors are able to explore such
questions whether the neutrino-nucleon cross section shows any
anomaly as compared to the Standard Model\cite{nestor}.

What is missing is that a typical underground or underwater
neutrino telescope cannot be easily made to be sensitive to
the primary neutrino energy or to explore the development
of a shower resulting from the interaction of the primary
neutrino. To be sure, there {\em are\/} some solutions
available to bridge the gap\cite{bianca}. Most of those are,
however, based on software development. It would be also
desirable to have detectors which are sensitive to the
primary neutrino energy directly. (The importance of this feature has been
emphasized in \mbox{ref.~(\cite{nestor}).)}

The basic idea we are here proposing is a very simple one.  UHE neutrinos 
originating either from a point source or from a diffuse background 
will interact in the Earth. If the interaction takes place sufficiently close 
to the surface of the Earth, at least a substantial part of the shower
originating from the primary interaction will develop in the atmosphere and
it can be viewed by OWL or, possibly, by a future detector mounted on the 
space station. Due to the fact that the interaction cross section of 
neutrinos is energy dependent, there is a one to one correspondence 
between the interaction mean free path (mfp) of the neutrino and its
impact parameter with respect to the center of the Earth. (Equivalently,
by an elementary exercise in geometry, the impact parameter dependence
can be  translated into a nadir angle dependence.)

The Earth acts as a filter: given a nadir angle and the corresponding mfp,
more energetic neutrinos will interact deep in the Earth and the shower
will die out before reaching the surface. Conversely, less energetic 
neutrinos either do not interact at all or they interact high in the 
atmosphere and there will be insufficient target thickness for the
shower to develop. It is to be emphasized, of course that this are
statements which hold for the average shower. Due to fluctuations
in the multiplicity of the first interaction and in the shower
development, one may not be able to determine shower energies 
on an event by event basis. (Nevertheless, simulations carried
out for the Fly's Eye detector indicate that even an event by
event energy determination may be feasible with a tolerable error.)
\section{An Example.} 
In order to determine the shower development under these assumptions,
our preliminary study concentrated upon the feasibility of observing
a neutrino induced shower. The question one has to decide is whether
the upward going shower develops over a substantial length in the
atmosphere so that it can be observed. (One has to recall that an
upward going shower starts in the dense part of the atmosphere,
near the Earth.) For this purpose, we calculated the longitudinal
development of showers in approximation A, assuming Feynman
scaling. It is known that Feynman scaling is violated in
hadronic interactions, due to QCD loop effects. The violation is, however,
logarithmical. We took the neutrino mfp from the work of 
Gandhi~\ETAL, ref.~\cite{gandhi}. An average hadronic cross
section of 60 mb was used, even though we find that the development of the
electromagnetic component does not depend very sensitively on variations of
the hadronic cross section within a reasonable range. The results were obtained
(as usual) in target depth, measured in $g/cm^{2}$. Conversion to distances
was obtained assuming an exponential atmosphere, of a scale height of
7.8 km.

In the following Figure, we show the result of a calculation for a neutrino
incident at a nadir angle of \DEG{80}. According to the neutrino mfp given in 
ref.~\cite{gandhi} and the specifications given above, this corresponds to
a neutrino energy of about $10^{9}$GeV. In calculating the 
electromagnetic component,
we assumed an initial hadron multiplicity of $\approx 50$ and an inelasticity
of $\approx 0.5$. The Figure contains the longitudinal profile of the
leptonic component induced by one hadron. The reason for this is that one
may contemplate other multiplicities and inelasticities. In that case, the
result is obtained by a simple rescaling of the ordinate. (This is a
property of the diffusion approximation used here; for the present purposes, 
it is adequate. However, in the case of more sophisticated models, this
scaling property is only approximately valid.)
 \leavevmode
 \begin{center}
\epsfxsize=300truemm
 \epsfbox[110 185 1034 540]{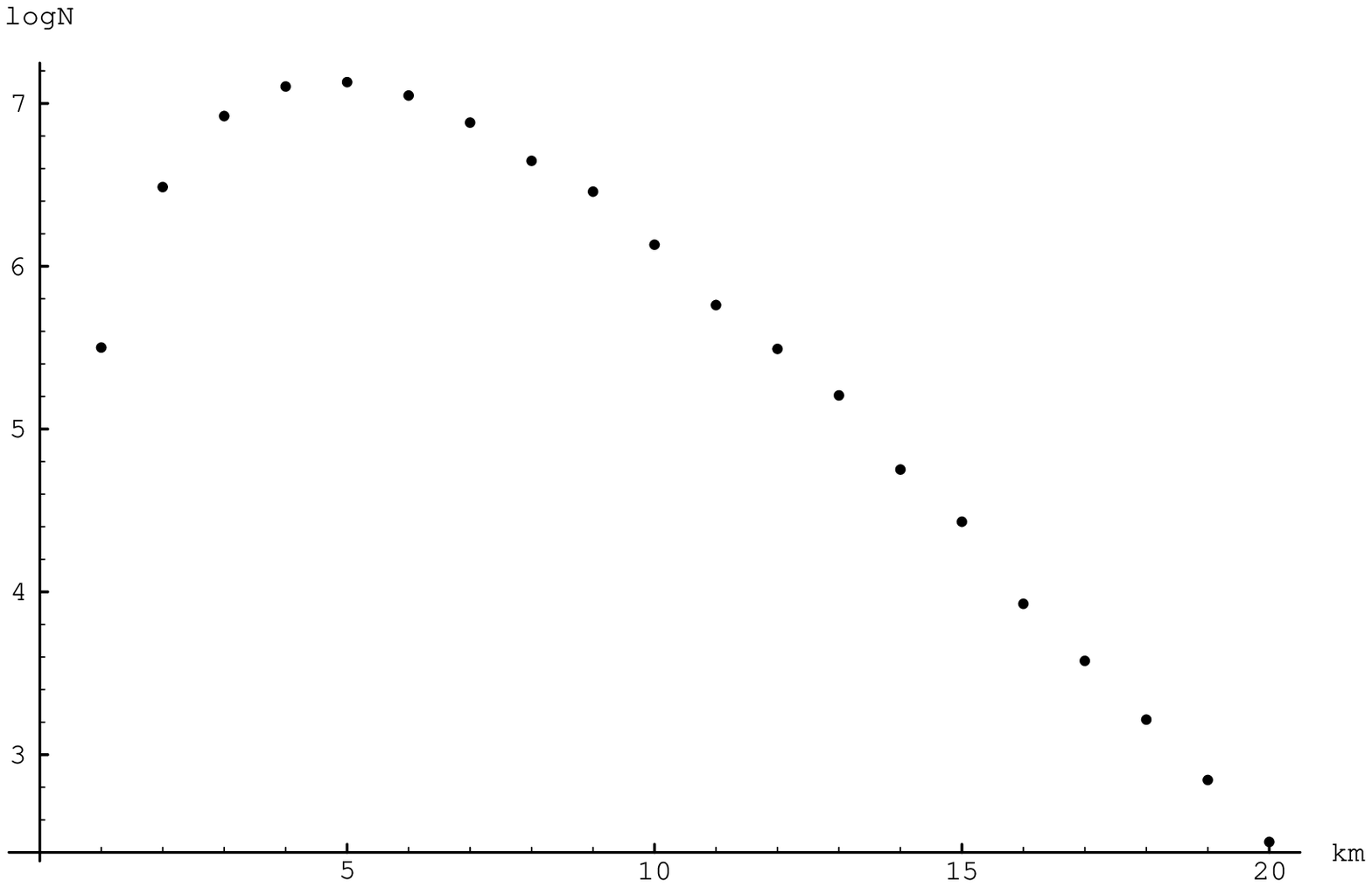}
Longitudinal profile of a shower described in the text.
\end{center}
\section{Discussion}
The example shown here suggests that one may be able to use orbiting
detectors in conjunction with the Earth as a target for the detection
of UHE neutrino interactions. At present, the properties of the detectors
are not yet sufficiently well known and thus, no quantitative conclusions
can be drawn. A more detailed calculation is needed in order to explore
particle physics capabilities of the orbiting neutrino telescopes. Work
along these lines is in progress.

We wish to thank the organizers of the Giant Air Shower meeting for a
very stimulating conference and Bianca Monteleoni for several useful
discussions on neutrino detection. 

\end{document}